\documentclass[prb,showpacs,altaffilletter,preprint]{revtex4}
\usepackage{graphicx}
\usepackage{dcolumn}
\usepackage{bm}
\usepackage{relsize}

\begin{document}
\newcommand{\FE}{\mathsmaller{FE}}
\title{Field-Effect Transistor on SrTiO${}_3$ with sputtered Al$\rm{}_2$O$\rm{}_3$ Gate Insulator}
\author{K.~Ueno}
 \altaffiliation[Electronic mail: ]{kazunori-ueno@aist.go.jp}
 \altaffiliation[Also at: ]{Department of Advanced Materials Science,
	University of Tokyo, Kashiwa, Chiba 277-8581, Japan}
 \affiliation{Correlated Electron Research Center (CERC), National
  Institute of Advanced Industrial Science and Technology (AIST),
  Tsukuba 305-8562, Japan}

\author{I.~H.~Inoue}
 \affiliation{Correlated Electron Research Center (CERC), National
  Institute of Advanced Industrial Science and Technology (AIST),
  Tsukuba 305-8562, Japan}

\author{H.~Akoh}
 \affiliation{Correlated Electron Research Center (CERC), National
  Institute of Advanced Industrial Science and Technology (AIST),
  Tsukuba 305-8562, Japan}

\author{M.~Kawasaki}
\altaffiliation[Also at: ]{Institute for Materials Research, Tohoku
  University Sendai 980-8577, Japan}
\affiliation{Correlated Electron Research Center (CERC), National
  Institute of Advanced Industrial Science and Technology (AIST),
  Tsukuba 305-8562, Japan}

\author{Y.~Tokura}
\altaffiliation[Also at: ]{Department of Applied Physics, University
  of Tokyo, Bunkyo-ku, Tokyo 113-8656, Japan}
\affiliation{Correlated Electron Research Center (CERC), National
  Institute of Advanced Industrial Science and Technology (AIST),
  Tsukuba 305-8562, Japan}

\author{H.~Takagi}
 \altaffiliation[Also at: ]{Department of Advanced Materials Science,
	University of Tokyo, Kashiwa, Chiba 277-8581, Japan}
 \affiliation{Correlated Electron Research Center (CERC), National
  Institute of Advanced Industrial Science and Technology (AIST),
  Tsukuba 305-8562, Japan}
 
\begin{abstract}
A field-effect transistor that employs a perovskite-type $\rm SrTiO_3$ single
crystal as the semiconducting channel is revealed to function as n-type
accumulation-mode device with characteristics similar to that of organic FET's.
The device was fabricated at room temperature by sputter-deposition of amorphous
$\rm Al_2O_3$ films as a gate insulator on the $\rm SrTiO_3$ substrate. 
The field-effect\,(FE) mobility is 0.1\,$\rm cm^2/Vs$ and on-off ratio exceeds 100
at room temperature.  The temperature dependence of the FE mobility down to 2\,K
shows a thermal-activation-type behavior with an activation energy of 0.6\,eV.
\end{abstract}
\date{\today}

\maketitle

Integrated circuits fully coordinated by functional oxide materials is still an eidolon of the next-generation electronic device.  A transition-metal oxide, $\rm SrTiO_3$, has been providing us an ideal laboratory for the fundamental investigation to such oxide electronics, being widely used as one of the most important bulk substrates.  It has a simple perovskite-type structure, and fairly smooth surface can be obtained through a chemical etching\cite{sto_sample}.  This atomically smooth surface plays a substantial role in realizing a perfect epitaxial growth of two-dimensional heterostructures --- a surpassing stage setting for the electronic-device application of the functional oxides.   Overture to such expedition is to fabricate a naive metal-insulator-semiconductor field-effect transistor\,(MISFET) and let it function sufficiently.  Nevertheless, although there  have been a couple of reports on the simple MISFET with a $\rm SrTiO_3$ channel\cite{sto_fet,italy_sto}, the on-off ratio for an applied field up to $\rm 1.5\,MV/cm$ was only about two, which is far insufficient for a switching device.

In this Letter, we report a new approach to the fabrication of oxide devices.  We fabricated an FET on a non-doped insulating $\rm SrTiO_3$ single crystal, and revealed that it shows an exemplary FET characteristics unprecedented in any FET's made of perovskite-type oxide semiconductors so far\cite{mott_fet}.  Three chisels have carved the niche: one is that we have used a surface of the bulk $\rm SrTiO_3$ single crystal as the channel. Another is that we have fabricated the FET structure by a so-called ``dry process'' with a stencil mask in order to prevent the surface of the $\rm SrTiO_3$ from being degraded by the standard lithographic technique. The third is that we have selected amorphous $\rm Al_2O_3$ films for the gate insulator which can be deposited on the $\rm SrTiO_3$ surface without significant damages while keeping a large breakdown strength as well as large capacitance.   However, with candid appreciation, our FET device\,[Fig.1(a)] is a reverse-engineering of that described in recent publications of Sch\"{o}n {\it et al.}, though the yet perplexing published results were declared to be fraudulent. Our work thus plays a role to give an insight into the feasibility of the claims as well as suggests essential requisites.

We used $\rm 10\times 10\times 0.5\,mm$ single-crystalline $\rm SrTiO_3$ pieces, which were polished and etched by a vendor\cite{shinkosha} according to Ref[1].  Atomic force microscopy images of the surface exhibited a clear step-and-terrace structure.  At first, source and drain electrodes of 20nm thickness aluminum were evaporated on the (100) surface of the $\rm SrTiO_3$ substrate from a resistively heated tungsten boat through a Ni stencil mask at a rate of $5\sim10\rm \,\AA/s$ under a pressure of $\rm 10^{-3} \,Pa$.  The aluminum electrodes for $\rm SrTiO_3$ is reported to be prefeable\cite{shimizu}. Actually, our electrodes yield an Ohmic contact on $\rm SrTiO_3$ as clearly seen in the inset of Fig.2.  It is worth mentioning that we kept a considerable distance ($\sim\rm 0.1\,mm$) between the stencil mask and the substrate during the evaporation in order to prevent the substrate from being contaminated by the mask. Then, the amorphous $\rm Al_2O_3$ films of 50nm thickness were prepared by a radio-frequency magnetron sputtering with a $\rm 99.99\,\%$\, pure ceramic $\rm Al_2O_3$ target ( 50\,mm$\phi$ ) in 2\,Pa flowing Ar gas.  The base pressure of the sputtering chamber was better than $\rm 7\times10^{-5}\,Pa$.  Sputtering power was 100\,W and the target-to-substrate\,(TS) distance was fixed at 120\,mm.  This TS distance seems to be preposterously large, but is a key point to realize a {\it gentle} deposition to the substrate.  Actually, the substrate temperature was at maximum 40\,$\rm{}^\circ C$ during the deposition, even though we do not equip any cooling systems. At the final stage of the process, a gold wire was attached on the top of the $\rm Al_2O_3$ film by conducting gold paint, which performs as the gate electrode, as shown in Fig.1\,(b).  Since our $\rm SrTiO_3$ substrate was quite insulating, it is important that the gate electrode, {\it i.e.} the gold paint, must cover the entire channel region. Otherwise, even a narrow channel region uncovered by the gate electrode might prevent a current flowing between the source and drain electrodes.

The average breakdown voltage of the deposited $\rm Al_2O_3$ film was 20\,V, which corresponds to 4\,$\rm MV/cm$, and the leakage current at the breakdown was several nA.  The capacitance per unit area $C_i$ was $0.16\,\mu\rm F/cm^2$.  Though we have attempted to optimize the sputtering condition over a wide range of deposition parameters, a maximum breakdown field has never exceeded $\rm 10MV/cm$.  The source-drain current-voltage\,($I_{DS}$-$V_{DS}$) characteristics with the $\rm Al_2O_3$/$\rm SrTiO_3$ FET with a channel length $L$ and a width $W$ of $\rm 25\,\mu m$ and $\rm 300\,\mu m$, respectively, were measured for different gate voltages $V_{GS}$.  The $I_{DS}$-$V_{GS}$ characteristics were measured for the temperature elevating from 2 to 400\,K by 1\,K or 2\,K step; the measurement for each temperature took about 5 minutes.  All measurements was carried out using the Agilent Technologies 4155C semiconductor parametric analyser.

The $\rm Al_2O_3$/$\rm SrTiO_3$ device shows a typical n-channel FET behavior as shown in Fig.2.  Application of a positive gate bias greatly enhanced $I_{DS}$, while $I_{DS}$ did not increase much for $V_{GS}$ below 2\,V.  ( No device we fabricated shows $I_{DS}$ enhancement by applying a negative gate bias until the breakdown.) This enhancement of $I_{DS}$ is due to the accumulation of negative carriers at the interface between the $\rm Al_2O_3$ insulator and the $\rm SrTiO_3$ substrate.  A saturation of $I_{DS}$ ({pinch-off}) was also observed for small values of $V_{DS}$.  This pinch-off behavior indicates that the channel region is sufficiently depleted in this $\rm Al_2O_3$ / $\rm SrTiO_3$ FET, while the behavior has been hardly observed in other ``dirty'' FET's such as that of amorphous Si.  An on-off ratio is one of the most important parameters of FET devices. It is a ratio of $I_{DS}$ for a given gate bias to that for zero gate bias.  The on-off ratio exceeds 100, which was deduced for $V_{GS}$ of 0\,V and 4\,V at $V_{DS}$ of 1\,V from the transfer characteristics shown in Fig.3\,(a).  This value is considerably larger than that of a similar device reported in literature\cite{italy_sto}, because their device was fabricated at much higher temperatures and therefore suffered from a large background current even in the off-state. The threshold voltage of 1.5\,V was also determined from Fig.3(a).

An interesting feature is that the field effect\,(FE) mobility $\mu_{\FE}$ of our $\rm Al_2O_3$/$\rm SrTiO_3$ FET depends on $V_{GS}$ up to the value as large as we could apply.  In conventional FET, $\mu_{\FE}$ can be well defined as
\[\mu_{\FE}^{lin} = \frac{\partial I_{DS}}{\partial V_{GS}}\left(\frac{L}{C_iWV_{DS}}\right)\ \ \ (1)\]
for large $V_{GS}$, {\it i.e.} the whole channel region is filled with the accumulated carriers. For $V_{GS}$ near the threshold value $V_{th}$, $\mu_{\FE}$ can be alternatively defined as
\[\mu_{\FE}^{sat} = \left(\frac{\partial \sqrt{I_{DS}}}{\partial V_{GS}}\right)^2\frac{2L}{C_iW} \ \ (2)\]
and the channel is in pinch-off state.  As shown in Fig.3 (b), both $\mu_{\FE}^{lin}$ and $\mu_{\FE}^{sat}$ increase monotonically with increasing $V_{GS}$ above $V_{th}$.  This large field dependence of the FE mobility is considered to be caused by the change of the field-induced carrier density; {\it i.e.}, the channel region changes drastically  from the insulating state to the more conducting state. Similar feature was reported for several accumulation-mode FET's on organics and amorphous-Si\cite{asi_gatedepmobility,org_gatedepmobility5}.

Since $\rm SrTiO_3$ shows an insulator to metal transition at very low carrier concentration\,($\sim 10^{18} \rm cm^{-3}$) and the mobility of metallic $\rm SrTiO_3$ increases with decreasing temperature,\cite{sto_hall} the temperature dependence of $\mu_{\FE}$ of our $\rm Al_2O_3$ / $\rm SrTiO_3$ FET for various gate biases is worth exploring.  When $V_{GS}$ is much larger than $V_{DS}$, $\mu_{\FE}^{lin}$ must be a good measure of the intrinsic $\mu_{\FE}$.  Therefore, we used $\mu_{\FE}^{lin}$ for $V_{GS}$ above 4\,V and $V_{DS}$ of 1\,V.  The device measured here has L of $\rm 100\,\mu m$ and W of $\rm 400\,\mu m$.  As shown in Fig.4, no metallic behavior was observed for $V_{GS}$ up to 9\,V.  ($V_{GS}$ of 9\,V corresponds to the field induced carrier density of $\rm 10^{13}\,cm^{-2}$.) Indeed, below 270K, $\mu_{\FE}^{lin}$ for such gate biases are dominated by the thermal activation, though from 180K to 2K  the signal was smaller than the noise level of $\rm 10^{-7}\,cm^2/Vs$.  The activation energy $E_a$ between 270\,K and 230\,K was deduced to be $\rm 0.6\,eV$ from the equation $\mu_{\FE}=\mu_0\exp(-E_a/k_{\rm B}T)$, where $k_{\rm B}$ is the Boltzmann constant.  This value of $E_a$ is independent of $V_{GS}$ between 4\,V and 9\,V and is much larger than the bulk value of 0.14\,eV, which was estimated by a time-of-flight technique\cite{stodrift_ins}.  Above 300\,K, the $\mu_{\FE}$ seems to be almost independent of $V_{GS}$, suggesting the transport mechanism was changed above room temperature.  At higher temperatures above 320K, since the gate leakage current increases rapidly with increasing temperature and becomes comparable to $I_{DS}$, it is not possible to reliably estimate the mobility.

The gate-dependent and thermal-activation-type mobility is often seen in the accumulation-mode FET of organics, cuprates and amorphous-Si\cite{org_gatedepmobility5,ybco_fet,asi_gatedepmobility}.  This is considered to reflect the nature of localized levels in the energy gap originated from dangling bonds or grain boundaries. In our $\rm Al_2O_3$/$\rm SrTiO_3$ device, the interface traps due to any surface defects or absorbed contaminants, which were inevitably introduced during the surface  preparation by the chemical etching, contributed to the behavior of $\mu_{\FE}$.  Indeed, there still needs a further investigation for the improvement of the device performance, among which the {\it in-situ} surface treatment by {\it e.g.} an annealing or a homoepitaxial deposition of a buffer layer between $\rm Al_2O_3$ and $\rm SrTiO_3$ should be considered.

In conclusion, we have demonstrated that an accumulation-mode MISFET devices with low threshold voltages and high on-off ratio can be realized using the surface of the bulk $\rm SrTiO_3$ single crystal as a channel and the sputter-deposited amorphous $\rm Al_2O_3$ as a gate insulator. The gentle deposition of the $\rm Al_2O_3$ reduced the off-state conductance drastically and the on-off ratio exceeds 100 by the applied gate electric fields of $\rm 0.8\,MV/cm$.  The FE mobility was found to be gate-bias dependent and thermal-activation-type as observed in other accumulation-mode FET's on several materials.  We expect that more elaborate interface preparation would improve the device performance.  However, the most important message we cast is that we have demonstrated the feasibility of the dry process with the sputtering of amorphous $\rm Al_2O_3$ directly on the top of any substrates. With this recipe, one can fabricate other FET devices on any functional materials which are to be epitaxially grown on the $\rm SrTiO_3$ substrates, and  a gate electric field enables to tune the electronic  properties of such materials --- the second stage of the exploration for future oxide electronics. 

We would like to thank A.~Sawa, H.~Sato, R.~Kumai, T.~Ito T.~Okuda, T.~Shimizu, T.~Yamada and Y.~Ishii for their help and valuable discussions.

\newpage

\begin{center}
 FIGURES
\end{center}

\noindent
Fig 1 :
(a) Schematic drawing of a field effect transistor\,(FET) structure fabricated on a single crystalline $\rm SrTiO_3$.  The field induced carrier is expected to be accumulated at the interface between the amorphous $\rm Al_2O_3$ insulator\,($\rm\sim 50nm$) and the $\rm SrTiO_3$ single crystal. 
(b) Photograph of the FET device on $\rm SrTiO_3$. Before the $\rm Al_2O_3$ deposition, a hillock of gold paint was heaped on each of the source\,(S) and drain\,(D) electrode. (Al electrodes are seen as dark squares.)  After the $\rm Al_2O_3$ deposition,  the top of the hillock was scraped off and attached to the lead wire.  The gate electrode\,(G) on the $\rm Al_2O_3$ layer is the gold paint covering a part of the gap between the source and drain electrodes. The active channel is  formed below the gate electrode underneath the $\rm Al_2O_3$ film. 

\vskip 20pt
\noindent
Fig 2 :
Drain-source current $I_{DS}$ plotted against the drain-source bias $V_{DS}$ of the $\rm Al_2O_3$/$\rm SrTiO_3$ FET at 300K.  A channel length and a width of the FET device were $\rm 25\,\mu m$ and $\rm 300\,\mu m$, respectively.  The inset shows the blow-up of the $I_{DS}$-$V_{DS}$ curve for $V_{GS}$\,=\,0\,V.

\vskip 20pt
\noindent
Fig 3 :
(a) The gate-source bias $V_{GS}$ dependence of the drain-source current $I_{DS}$ for a fixed drain-source bias $V_{DS}$ = $\rm +1\,V$ of the same device used for Fig.2.  The on-off ratio between $V_{GS}$ of 0\,V and 4\,V for $V_{DS}$ of 1\,V exceeds 100.
(b) $V_{GS}$ dependence of the field effect mobility $\mu_{\FE}$. $\mu_{\FE}^{lin}$ and $\mu_{\FE}^{sat}$ were deduced from Fig.3\,(a) by using Eq\,(1) and Eq\,(2), respectively.  Both increases monotonically with $V_{GS}$ and no saturation was observed even for large gate bias.

\vskip 20pt
\noindent
Fig 4 :
Field-effect mobility $\mu_{\FE}$ as a function of temperature. $\mu_{\FE}$ was deduced by using Eq.(1) for the gate-source bias $V_{GS}$ of 4\,V, 6\,V, 9\,V.  The device has a channel length of $\rm 100\,\mu m$ and a width of $\rm 400\,\mu m$.  The solid line corresponds to a thermally activated behavior with an activation energy $E_a$ of 0.6\,eV.

\newpage
\begin{flushleft}
\includegraphics[width=15cm]{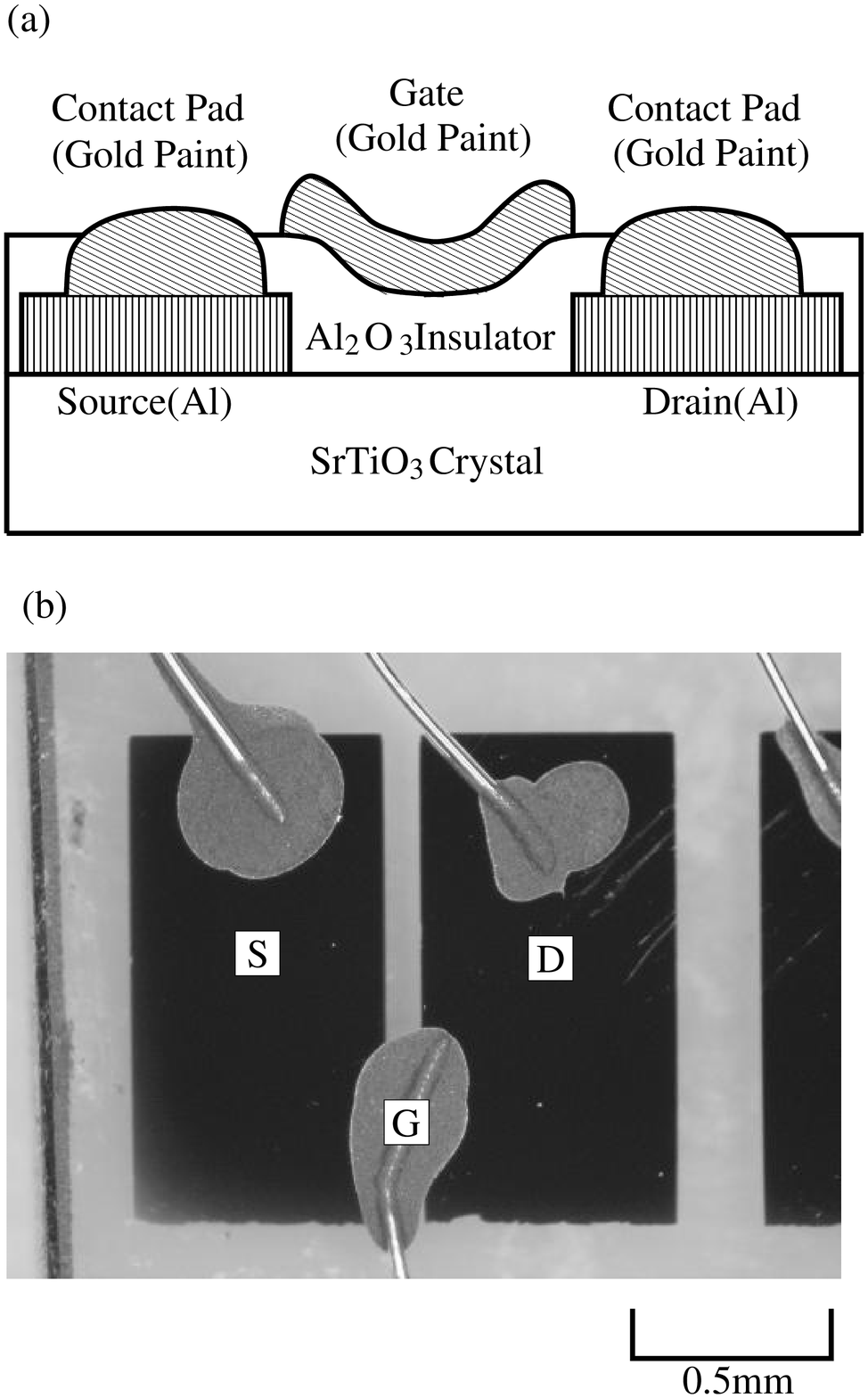}
\end{flushleft}

\newpage
\begin{flushleft}
\includegraphics[width=15cm]{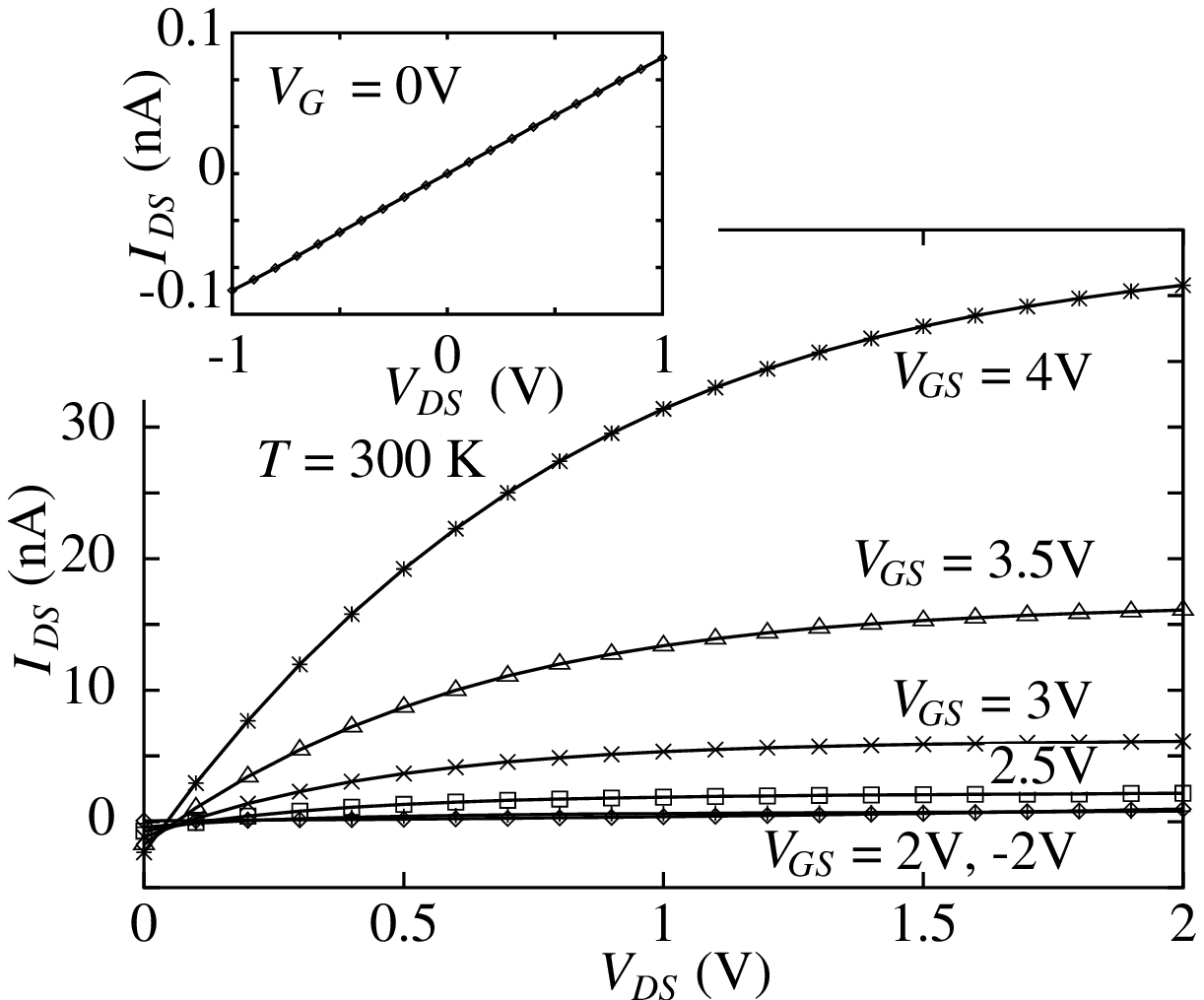}
\end{flushleft}

\newpage
\begin{flushleft}
\includegraphics[width=15cm]{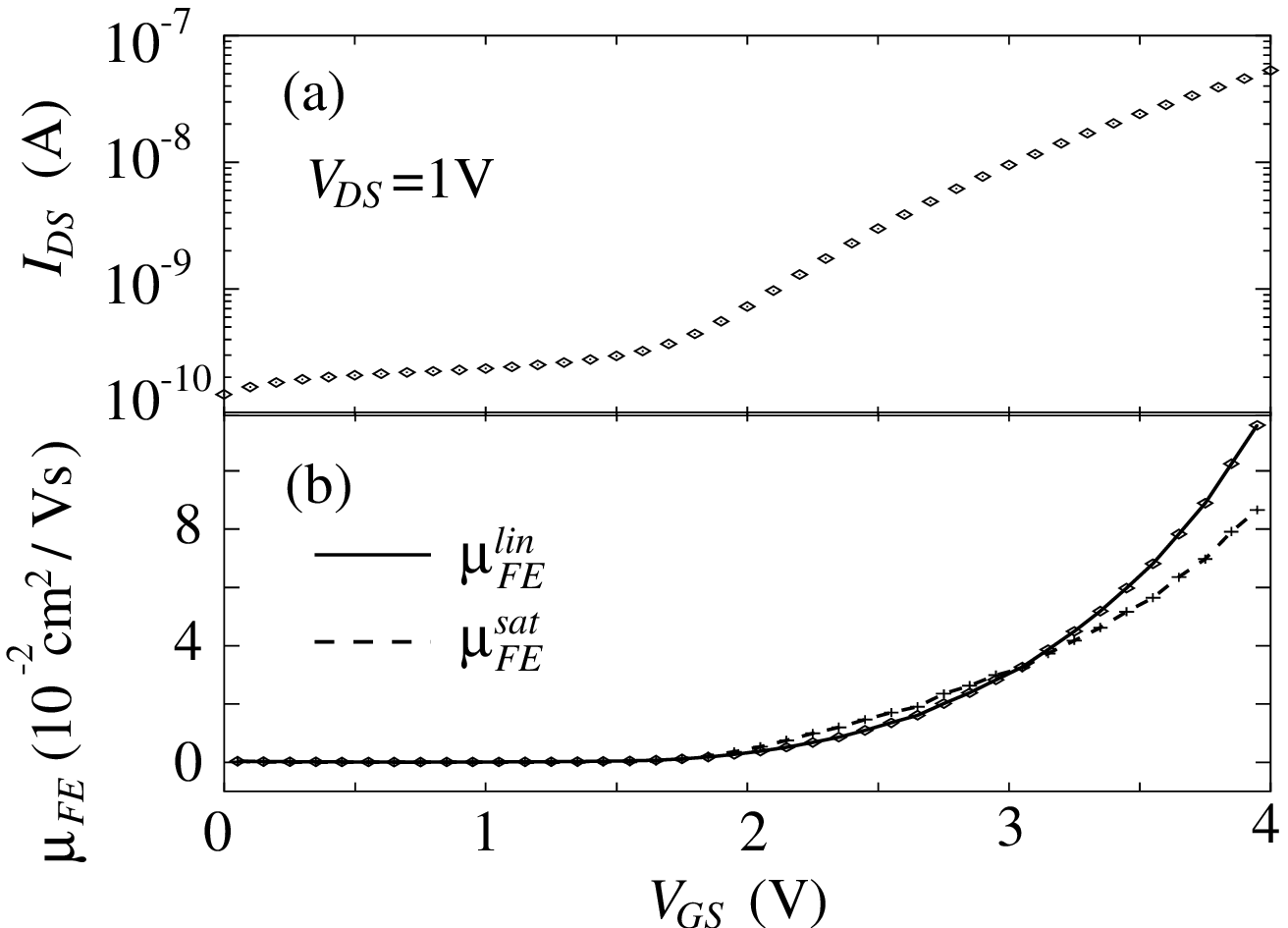}
\end{flushleft}

\newpage
\begin{flushleft}
\includegraphics[width=15cm]{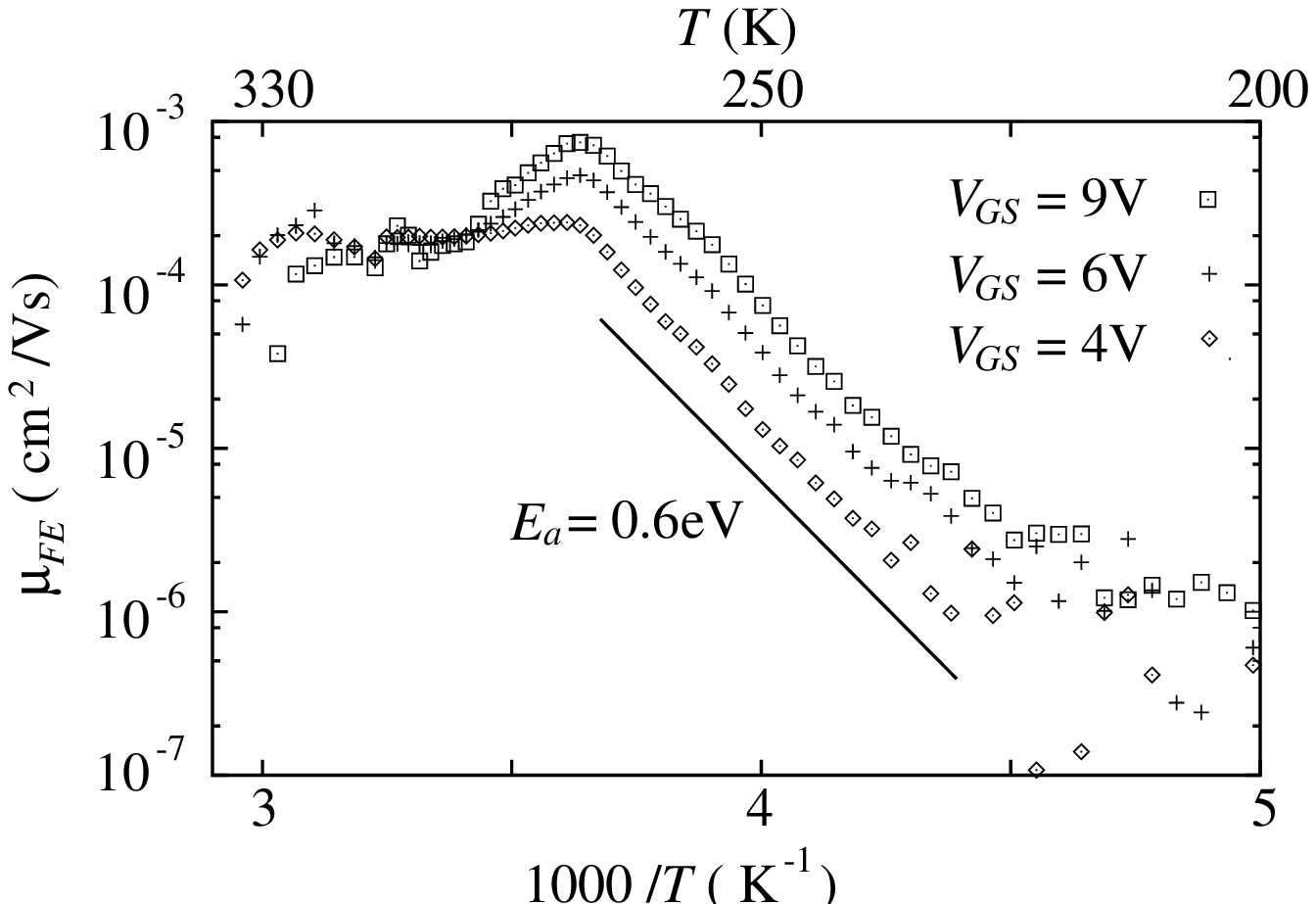}
\end{flushleft}

\end{document}